\pdfoutput=1
\documentclass[journal]{IEEEtran}

\usepackage[english]{babel}
\usepackage{graphicx,url}       
\usepackage{mathptmx}           
\usepackage{epsfig}
\usepackage{subfigure}
\usepackage[dvipsnames,svgnames,table]{xcolor}
\usepackage{hhline}
\usepackage{etoolbox}
\usepackage{cellspace}
\usepackage{adjustbox}

\usepackage{pdfpages}
\usepackage{url}
\usepackage{color}
\usepackage{xspace}

\usepackage{listings}
\usepackage{tabularx}
\usepackage{array}
\usepackage{pbox}

\usepackage{multirow} 
\usepackage{setspace}
\usepackage{amsmath}
\usepackage{amssymb}
\usepackage{amsfonts}
\usepackage{amsthm}

\usepackage[utf8]{inputenc}
\usepackage[T1]{fontenc}
\usepackage{algpseudocode,algorithm,algorithmicx}
\definecolor{MyBlue}{HTML}{5f8dd3}

\definecolor{lightgray}{gray}{0.9}
\definecolor{darkgray}{gray}{0.5}

\usepackage[nolist,nohyperlinks]{acronym}

\usepackage{hyperref}
\hypersetup{
  draft=false,
  colorlinks=true,
  linktocpage=true,
  pdfstartview=FitH,
  breaklinks=true,
  pdfpagemode=UseNone,
  pageanchor=true,
  pdfpagemode=UseOutlines,
  plainpages=false,
  bookmarksnumbered,
  bookmarksopen=false,
  bookmarksopenlevel=1,
  hypertexnames=true,
  pdfhighlight=/O,
  urlcolor=Maroon,linkcolor=MyBlue,citecolor=DarkGreen, 
  pdftitle={},
  pdfauthor={},
  pdfsubject={},
  pdfkeywords={},
  pdfcreator={pdfLaTeX},
  pdfproducer={LaTeX with hyperref}
}

\usepackage[textwidth=30mm]{todonotes}
\usepackage{soul}
\setstcolor{red}
\sethlcolor{SkyBlue}

\theoremstyle{plain}

\newtheorem*{corollary*}{Corollary}

\theoremstyle{definition}

\newtheorem*{definition*}{Definition}

\theoremstyle{remark}

\newtheorem*{remark*}{Remark}

\newcommand{\flowset}{\mathcal{S}}

\newcommand{\flowid}{k}

\newcommand{\flowidset}{\mathcal{K}}

\newcommand{\feature}{a}
\newcommand{\featureset}{\Omega}

\newcommand{\nodei}{i}
\newcommand{\nodej}{j}

\newcommand{\alphabet}{\Omega}

\newcommand{\field}{h}

\newcommand{\featurei}{i}
\newcommand{\featurej}{j}

\newcommand{\featureids}{\mathrm{N}}

\newcommand{\coup}{e}

\newcommand{\edge}{\varepsilon}
\newcommand{\func}{G(\node,\edge)}
\newcommand{\funck}{G_\flowid(\node,\edge)}

\newcommand{\spin}{a}

\newcommand{\ie}{\textit{i.e., }}                   
\newcommand{\eg}{\textit{e.g., }}                   
\newcommand{\etal}{\textit{et al. }}               

\newcommand{\node}{\eta}

\begin{acronym}
  \acro{ANN}{Artificial Neural Network}
  \acro{MLP}{Multi-Layer Perceptron}
  \acro{NIDS}{Network Intrusion Detection System}
  \acro{NID}{Network Intrusion Detection}
  \acro{EFC}{Energy-based Flow Classifier}
  \acro{DoS}{Denial of Service}
  \acro{ML}{Machine Learning}
  \acro{DT}{Decision Tree}
  \acro{LR}{Linear Regression}
  \acro{NB}{Naive Bayes}
  \acro{KNN}{K-Nearest Neighbors}
  \acro{RF}{Random Forest}
  \acro{SVM}{Support Vector Machine}
  \acro{IoT}{Internet of Things}
  \acro{SDN}{Software Defined Networks}
  \acro{CNN}{Convolutional Neural Network}
  \acro{PCA}{Principal Component Analysis}
  \acro{DBN}{Deep Belief Network}
\end{acronym}

\begin{document}
\title{A new method for flow-based network intrusion detection using the inverse Potts model}

\author{Camila~Pontes,
        Manuela~Souza,
        Jo\~ao~Gondim,
        Matt~Bishop
        and~Marcelo~Marotta
\thanks{This manuscript was received at 15 of october and accepted 15 of march - Matt Bishop was supported by the  National  Science  Foundation  under  Grant  Number  OAC-1739025.  Any  opinions,  findings,  and  conclusions  or  recommendations expressed in this material are those of the author(s)and do not necessarily reflect the views of the National Science Foundation. Collaborative: Securing Networks in the Programmable Data Plane Era” funded by NSF (National Science Foundation), RNP (Brazilian National Research Network) and GigaCandanga. The associate editor coordinating the review of this article and approving it for publication was Q. Li. (Corresponding author: Camila F. T. Pontes.)}
\thanks{C. Pontes, M. Souza, J. Gondim and M. A. Marotta are with the University of Brasilia, Brazil, emails: cftpontes@gmail.com, gondim@unb.br, marcelo.marotta@unb.br;}
\thanks{M. Bishop is with the University of California at Davis, Davis, USA, email: mabishop@ucdavis.edu}
\thanks{DOI: 10.1109/TNSM.2021.3075503}
}
 

\maketitle

\begin{abstract}
Network Intrusion Detection Systems (NIDS) play an important role as tools for identifying potential network threats.
In the context of ever-increasing traffic volume on computer networks, flow-based NIDS arise as good solutions for real-time traffic classification.
In recent years, different flow-based classifiers have been proposed using Machine Learning (ML) algorithms.
Nevertheless, classical ML-based classifiers have some limitations.
For instance, they require large amounts of labeled data for training, which might be difficult to obtain.
Additionally, most ML-based classifiers are not capable of domain adaptation, \ie after being trained on an specific data distribution, they are not general enough to be applied to other related data distributions.
And, finally, many of the models inferred by these algorithms are black boxes, which do not provide explainable results.
To overcome these limitations, we propose a new algorithm, called Energy-based Flow Classifier (EFC).
This anomaly-based classifier uses inverse statistics to infer a statistical model based on labeled benign examples.
We show that EFC is capable of accurately performing binary flow classification and is more adaptable to different data distributions than classical ML-based classifiers.
Given the positive results obtained on three different datasets (CIDDS-001, CICIDS17 and CICDDoS19), we consider EFC to be a promising algorithm to perform robust flow-based traffic classification.
\end{abstract}


\IEEEpeerreviewmaketitle

\section{Introduction}

\IEEEPARstart{S}{ymantec's} Internet Security Threat Report \cite{symantec_internet_2019} points out a 56\% increase in the number of web attacks in 2019. Network scans, denial of service, and brute force attacks are among the most common threats. Such malicious activities threaten individuals and collective organizations such as public health, financial, and government institutions. In this context, \acp{NIDS} play an important role as tools for managing and identifying potential threats in the network \cite{pradhan2020intrusion}.

There are two main approaches for \acp{NIDS} regarding the kind of data analyzed: packet-based and flow-based. 
In the former, deep packet inspection is performed, taking into account individual packet payloads and header information \cite{ring2019survey}.
In the latter, flows, \ie packet collections, are analyzed regarding their properties, \eg duration, number of packets, number of bytes, and source/destination port \cite{ring2019survey}. 
To perform classification in real-time, a massive volume of data must be analyzed, making deep packet inspection too costly to be applied regarding processing and energy consumption.
Since flow-based approaches can classify the whole traffic inspecting an equivalent to 0.1\% of the total volume, \acp{NIDS} based on flow analysis arise as good solutions for real-time traffic classification \cite{Sperotto2010}. Besides, with the advent of software-defined networking and the virtualization of network functions, distributed security systems can take advantage of the spread of  \acp{NIDS} based on flow analysis to improve their security management across the network~\cite{Chica2020}.

In recent years, different flow-based classifiers have been proposed based on both shallow and deep learning \cite{Umer2017}.
According to the report in \cite{Umer2017}, the best flow-based classifiers achieve around 99\% accuracy.
Although quite accurate, classical \ac{ML}-based classifiers require labeled malicious traffic samples to perform training. However, real traffic labeling might be difficult, especially in the case of malicious traffic \cite{singla2019overcoming}.
Besides, \ac{ML}-based classifiers after trained on specific data distribution usually do not work well when applied to other data with slightly different distributions, \ie they have low domain adaptation capability~\cite{bartos2016optimized, li2019dart, zolanvari2018effect}.
Such a capability is particularly important to the network context since a standard procedure is to perform the training of classifiers with simulated data and, afterward, apply in real scenarios that change the data distribution requiring domain adaptation.
Moreover, most \ac{ML} algorithms are well-known to be black-box mechanisms, challenging to be understood and readjusted in detail ~\cite{rudin2019stop, holzinger2018machine}.
In this regard, there is a clear need for a new flow-based classifier for \acp{NIDS}, which generates an understandable model (white box) based solely on benign examples, and adaptable to different domains. 

In this work, we propose a novel classifier called \ac{EFC}, which is inspired by the inverse Potts model from quantum mechanics and adapted to network flow classification.
\ac{EFC} performs one-class, anomaly-based classification, \ie as long as it can learn the properties of benign flows, it will discriminate between benign and malicious flows.
Moreover, it is a white box algorithm, producing a statistical model that can be analyzed in detail regarding individual parameter values.
Here, we compared the performance of \ac{EFC} against a variety of classifiers using three different datasets, \ie CIDDS-001 \cite{ring2017flow}, CICIDS17 \cite{sharafaldin2018toward}, and  CICDDoS19 \cite{sharafaldin2019developing}.
Our results show that EFC's performance is comparable to the performance of the other classifiers.
Also, we observed that EFC is less sensitive to changes in data distribution than the others.
Our main contributions are:
\begin{itemize}
    \item The proposal and implementation of a flow classifier based on the inverse Potts model to be employed in \acp{NIDS};
    \item A performance comparison of the proposed classifier with classical \ac{ML}-based classifiers using three different datasets;
    \item An analysis of how different classifiers perform when trained within one domain and tested in another related domain.
\end{itemize}

The rest of this paper is structured as follows. In Section~\ref{sec:rel_work}, we briefly present the state-of-the-art in flow-based \acp{NIDS}. In Section~\ref{sec:preliminaries}, we describe the structure of network flows with a preliminary analysis of the datasets considered here. In Section~\ref{sec:proposal}, we introduce the statistical model proposed and the classifier implementation. In Section~\ref{sec:results}, we present the results obtained regarding the statistical model's analysis and the classification experiments performed. Finally, in Section~\ref{sec:conclusions}, we present our conclusions and future work.

\section{Related Work} \label{sec:rel_work}
In this section, we first briefly review the state-of-the-art in flow-based network intrusion detection systems.
In the following, some previous work on CIDDS-001, CICIDS17, and CICDDoS19 are shown to highlight their relevance as up-to-date datasets to be used in our experiments.
Finally, some challenges of \ac{ML}-based traffic classification are regarded, such as the difficulty in obtaining sufficient labeled data, the non-interpretability of models, and the difficulty in adapting to different domains (data distributions).

Several \ac{ML}-based classifiers have been explored over the last years for network intrusion detection.
Vinayakumar \etal (2017) \cite{vinayakumar2017evaluating}, Mahfouz \etal (2020) \cite{mahfouz2020comparative} and Khan \etal (2020) \cite{khan2020performance} independently evaluated the performance of different ML-based classifiers over internet traffic datasets.
In \cite{vinayakumar2017evaluating}, the KDDCup'99 and NSL-KDD datasets are regarded to evaluate the performance of both shallow and deep learning-based classifiers.
It is shown that deep learning-based approaches performed better to differentiate malicious attacks from benign traffic.
Meanwhile, the authors of \cite{mahfouz2020comparative} considered the NSL-KDD dataset to compare the performance of different shallow learning-based classifiers.
The classifier that presented the best performance without feature selection was the \ac{DT}.
With feature selection, \ac{KNN} performed better to classify malicious traffic.
Finally, the authors of \cite{khan2020performance} compared the performance of a few different classifiers over the UNSW-NB15 dataset. They observed that \ac{RF} overperformed all other classifiers. In fact, \ac{RF} has been used in several recent \acp{NIDS} \cite{tan2019wireless, kazemitabar2019novel, bhavani2020network}.
All of the aforementioned works use the F1-score as a metric to assess the performance of the different classifiers. 
In the present work, we consider both deep and shallow learning-based classifiers as baselines to assess EFC's performance over three different datasets, regarding the F1-score as one of the evaluation metrics.

To assess \ac{EFC}'s performance, one of the datasets we use is CIDDS-001.
This dataset was used by Verma and Ranga (2018) \cite{verma2018statistical} to assess the performance of \ac{KNN} and k-means clustering algorithms. Both algorithms achieved over 99\% accuracy. Also, Ring \etal \cite{ring2018detection} explored slow port scans detection using CIDDS-001. The approach proposed by them is capable of accurately recognizing the attacks with a low false alarm rate. Finally, Abdulhammed \etal \cite{Abdulhammed2019} also performed classification based on flows on CIDDS-001 and proposed an approach that is robust considering imbalanced network traffic. In summary, CIDDS-001 is an updated and relevant dataset to be used for network flow-classification solutions, being one of our dataset choices for assessing the performance of EFC.

The other two datasets considered here are CICIDS17 and CICDDoS19, from the Canadian Institute for Cyber Security.
Recently, Yulianto, Sukarno, and Suwastika~\cite{yulianto2019improving} used CICIDS17 to assess the performance of an Adaboost-based classifier.
Aksu \etal~\cite{aksu2018intrusion} did the same in 2018 with different \ac{ML} classifiers.
CICIDS17 contains benign as well as the most up-to-date common attacks, resembling true real-world data, being a relevant dataset to consider for flow-based traffic classification.
Meanwhile, CICDDoS19 is a very recent dataset with a focus on DDoS attacks. \cite{li2020rtvd} proposes a real-time entropy-based \ac{NIDS} for detection of volumetric DDoS in \ac{IoT} and performs tests over CICDDoS19 dataset, among other datasets. Another recent work \cite{jia2020flowguard} obtained over 99\% accuracy over CICDDoS19 dataset using a \ac{CNN}. And, finally, Novaes \textit{et al.} \cite{novaes2020long} proposed a system for intrusion detection based on fuzzy logic, which had its performance assessed on CICDDoS19. The rising popularity of this dataset serves as evidence of its relevance to assess the performance of different \ac{NIDS}. Hence, we use CICIDS17 and CICDDoS19 datasets to test our classifier and compare it to the performance of classical \ac{ML} classifiers.

Umer et al. (2017) \cite{Umer2017} performed a comprehensive literature survey on flow-based network intrusion detection.
In their work, some disadvantages of using \ac{ML}-based classifiers for traffic classification are mentioned.
Among them are the high computational cost of training the classifiers, the difficulty in obtaining representative datasets, and the high false positive rates observed.
The present work addresses some of the issues mentioned, since the classifier proposed here has a low computational cost and learns exclusively based on benign samples.
In the following, these and some other issues are discussed in further detail.

One of the commonly discussed issues in the field of \ac{ML} is the tight dependency most algorithms have on the amount of labeled samples available for training \cite{singla2019overcoming}, which might be difficult to obtain in some contexts.
For instance, it is difficult to obtain malicious traffic samples and to label it in the real world and this is why most of the network intrusion detection datasets contain simulated attacks.
This issue makes it difficult to train intrusion detection algorithms in such a way that they might be able to detect zero-day threats \cite{singla2019overcoming}.
The only way of possibly detecting a zero-day attack is relying on an anomaly-based classifier \cite{6542524}, such as the one we propose in this work.
EFC, has a great advantage over other \ac{ML}-based algorithms, which is the capability to infer a model based solely on benign traffic samples, \ie half of the information.
Such capability can be used to circumvent the problem of obtaining a high amount of data and the labeling of malicious samples.

Another common problem in \ac{ML} is that inferred models lose their predictive performance when tested in different domains (data distributions) \cite{zolanvari2018effect}.
In the field of network security, this adaptability is specially important given the existence of zero-day threats and the artificiality of most datasets used for research.
In \cite{zolanvari2018effect}, there is an interesting discussion about the existing differences between datasets used by academics to test \acp{NIDS} and the network traffic observed in the real world.
Additionally, the work of Bartos \textit{et al.} \cite{bartos2016optimized} and Li \textit{et al.} \cite{li2019dart} also address this issue.
They propose similar approaches, applying data transformations to the data to reduce differences between data distributions in different domains.
In our work, we propose a classifier that is intrinsically adaptable to different domains, since the model inference is based solely on benign samples.
Therefore, there is no need to transform the data in order to adapt the model or perform adjustments to a different domain, making our approach simpler and more straightforward.

Finally, another big issue in \ac{ML} is the non-interpretability of some models \cite{rudin2019stop, holzinger2018machine}.
\acp{ANN}, in special, became more and more opaque with time, despite overperforming other approaches in many tasks. The authors of \cite{holzinger2018machine} highlight that the best \ac{ML} algorithms are not interpretable, hence the decision taken by them can not be explained.
However, different contexts require transparent decision making and that is why the development of explainable models is so important.
The authors of \cite{rudin2019stop} call attention to the fact that trying to explain black box models might not be the best approach to solve the issue of non-interpretability.
Instead, it is suggested the design of new models that are inherently interpretable.
In line with what has been suggested by these recent studies, EFC generates a white-box model and, therefore, satisfies the requirement of providing explainable results, allowing classification results to be analysed in retrospect if needed. Thus, next, we introduce main concepts and intuitions to serve as basis for EFC.

\section{Background and Datasets} \label{sec:preliminaries}
In this section, we present some fundamental concepts to understand flow-based network intrusion detection.
First, the concept of network flow and its features are introduced.
In the following, the three internet flow datasets used in this work are presented and described in detail to provide concrete examples of features and contextualize the experimental results presented in Section~\ref{sec:results}.
The information provided in this section serves as a basis to understand how EFC works.

A network flow is a set of packets that traverses intermediary nodes between end-points within a given time interval.
Under the perspective of an intermediary node, \ie an observation point, all packets belonging to a given flow have a set of common features called flow keys.
It means that flow keys do not change for packets belonging to the same flow, while the remaining features might vary.
FlowScan \cite{plonka2000flowscan} is an example of a tool capable of collecting data from a set of packets and extracting flow features to be later exported in different formats, such as NetFlow and IPFIX.
Since NetFlow is the most commonly used format, its main features are listed below:
    \begin{itemize}
        \item Source/Destination IP (flow keys) - determine the origin and destination of a given flow in the network;
        \item Source/Destination port (flow keys) - characterize different kinds of network services \eg ssh service uses port 22;
        \item Protocol (flow key) - characterizes flows regarding the transport protocol used \eg TCP, UDP, ICMP.
        \item Number of packets (feature) - total number of packets captured in a flow;
        \item Number of bytes (feature) - total number of bytes in a flow;
        \item Duration (feature) - total duration of a flow in seconds;
        \item Initial timestamp (feature) - system time when one flow started to be captured.
    \end{itemize}
Other features such as TCP Flags and Type of Service might also be exported in some cases. 
The combination of different flow keys and features characterize one flow and determine its particular behavior.

Flow-based approaches are seen as suitable alternatives to precede packet inspection in real-time \acp{NIDS}.    
The idea is to deeply inspect only the packets belonging to flows considered to be suspicious by the flow-based classifier. 
A two-step approach would notably reduce the amount of data analyzed while maintaining a high classification accuracy \cite{Sperotto2010}.
In this work, we are only concerned with the first step, which is the flow classification.
We evaluate the performance of our algorithm, the EFC, compared to other ML algorithms using three different datasets.
We also evaluate the performance of the algorithms by training with data from part of the dataset and testing with other parts of it. Nonetheless, although both parts of the data come from the same dataset, their distributions are different to characterize domain adaptation.
In the following, we briefly describe the datasets used for testing and characterize what constitutes a domain adaptation in each of them.

\subsection{CIDDS-001}
CIDDS-001 \cite{ring2017flow} is a relatively recent dataset composed of a set of flow samples captured within a simulated OpenStack environment and another set of flow samples obtained from a real server.
The former contains only simulated traffic, while the latter includes both real and simulated traffic.
Each sample collected within these two environments has one of the labels described in Table \ref{tab:labels-cidds001}.

\begin{table}[!h]
\renewcommand{\arraystretch}{1.3}
\caption{Labels within CIDDS-001 dataset}
\label{tab:labels-cidds001}
\centering
\begin{adjustbox}{max width=\columnwidth}
\begin{tabular}{ l l }
\hline
\textbf{Environment} & \textbf{Labels} \\
\hline
OpenStack & normal, DoS, portScan, pingScan, bruteForce \\
External server & normal, DoS, bruteForce, unknown, suspicious \\
\hline
\end{tabular} 
\end{adjustbox}
\end{table}

Simulated benign flows are labeled as \textit{normal}, while simulated malicious flows are labeled as \textit{dos}, \textit{portScan}, \textit{pingScan} or \textit{bruteForce}, depending on the type of attack simulated.
The labels \textit{suspicious} and \textit{unknown}, in turn, are used for real traffic.
The external server is open to user access through the ports 80 and 443.
Hence, flows directed at these ports were labeled as \textit{unknown}, since they could be either benign or malicious.
All flows directed at other ports were labeled as \textit{suspicious}.
Traffic was sampled in both the simulated and the external server environment for a period of four weeks.
Within this dataset, a change from the simulated data distribution to the external server data distribution is a domain change, requiring the classifiers to adapt.

CIDDS-001 dataset flow features are shown in Table~\ref{tab:features}.
All features were taken into account for characterization and classification except for \textit{Src IP}, \textit{Dest IP} and \textit{Date first seen}.
These exceptions are because the latter one is intrinsically not informative to differentiate flows, and the former two are made up in the context of the simulated network and might be confounding.

\begin{table}[!h]
\renewcommand{\arraystretch}{1.3}
\caption{Features within CIDDS-001 dataset}
\label{tab:features}
\centering
\begin{adjustbox}{max width=\columnwidth}
\begin{tabular}{ l l l }
\hline
\textbf{\#} & \textbf{Name} & \textbf{Description} \\
\hline
1 & Src IP & Source IP Address \\
2 & Src Port & Source Port \\
3 & Dest IP & Destination IP Address \\
4 & Dest Port & Destination Port \\
5 & Proto & Transport Protocol (\eg  ICMP, TCP, or UDP) \\
6 & Date first seen & Start time flow first seen \\
7 & Duration & Duration of the flow \\
8 & Bytes & Number of transmitted bytes \\
9 & Packets & Number of transmitted packets \\
10 & Flags & OR concatenation of all TCP Flags \\
\hline
\end{tabular}
\end{adjustbox}
\end{table}

\subsection{CICIDS17}
CICIDS17 \cite{sharafaldin2018toward} dataset contains benign traffic and the most up-to-date common attacks, resembling real-world data.
This dataset was built using the abstract behavior of 25 users based on the HTTP, HTTPS, FTP, SSH, and email protocols.
The data was captured during one week in July 2017.
The attacks implemented include Brute Force FTP, Brute Force SSH, DoS, Heartbleed, Web Attack, Infiltration, Botnet, and DDoS.
They were executed both morning and afternoon on Tuesday, Wednesday, Thursday, and Friday (see Table \ref{tab:attacks-cicids}).

\begin{table}[!h]
\renewcommand{\arraystretch}{1.3}
\caption{Attacks within CICIDS17 dataset}
\label{tab:attacks-cicids}
\centering
\begin{adjustbox}{max width=\columnwidth}
\begin{tabular}{ l l }
\hline
\textbf{Week day} & \textbf{Attacks} \\
\hline
Monday &  \\
Tuesday & FTP-Patator, SSH-Patator \\
Wednesday & \begin{tabular}[x]{@{}l@{}} DoS slowloris, DoS Slowhttptes, DoS Hulk, DoS \\GoldenEye, Heartbleed Port 444 \end{tabular}  \\
Thursday & \begin{tabular}[x]{@{}l@{}} Brute Force, XSS, Sql Injection, Dropbox download, \\ Cool disk \end{tabular} \\
Friday & Botnet ARES, Port Scan, DDoS LOIT \\
\hline
\end{tabular}
\end{adjustbox}
\end{table}

Flow features on this dataset were extracted using CICFlowMeter \cite{lashkari2017characterization}.
There are in total 88 features, which are not going to be cited here because of the limited space.
All features were considered here, except for Flow ID, Source IP, Destination IP, and Timestamp.
These exceptions were made because the features were either intrinsically not informative or made up within a simulated environment.

\subsection{CICDDoS19}
CICDDoS19 \cite{sharafaldin2019developing} contains benign traffic and the most up-to-date common DDoS attacks (volumetric and application: low volume, slow rate), resembling real-world data. This dataset contains different modern reflective DDoS attacks such as PortMap, NetBIOS, LDAP, MSSQL, UDP, UDP-Lag, SYN, NTP, DNS, and SNMP.
The traffic was captured in January (first day) and March (second day), 2019.
Attacks were executed during this period (see Table \ref{tab:attacks-cicddos}).

\begin{table}[!h]
\renewcommand{\arraystretch}{1.3}
\caption{Attacks within CICDDoS19 dataset}
\label{tab:attacks-cicddos}
\centering
\begin{adjustbox}{max width=\columnwidth}
\begin{tabular}{ l l }
\hline
\textbf{Day} & \textbf{Attacks} \\
\hline
First & PortMap, NetBIOS, LDAP, MSSQL, UDP, UDP-Lag, SYN \\
Second & \begin{tabular}[x]{@{}l@{}} NTP, DNS, LDAP, MSSQL, NetBIOS, SNMP, SSDP, UDP, \\UDP-Lag, WebDDos, SYN, TFTP \end{tabular}  \\
\hline
\end{tabular} \vspace{0pt}
\end{adjustbox}
\end{table}

Flow features on this dataset were extracted using CICFlowMeter \cite{lashkari2017characterization}.
All features were considered here, except for Flow Id, Source IP, Destination IP, and Timestamp. These exceptions were made because the features were either intrinsically not informative or made up within a simulated environment. 

Considering each concept regarding network flows, their features, and how they are presented across different datasets, serve as basis to introduce the main intuition behind EFC, such as presented next.

\section{Proposal} \label{sec:proposal}
\begin{figure}[H]
\centering
\includegraphics[width=\columnwidth, clip]{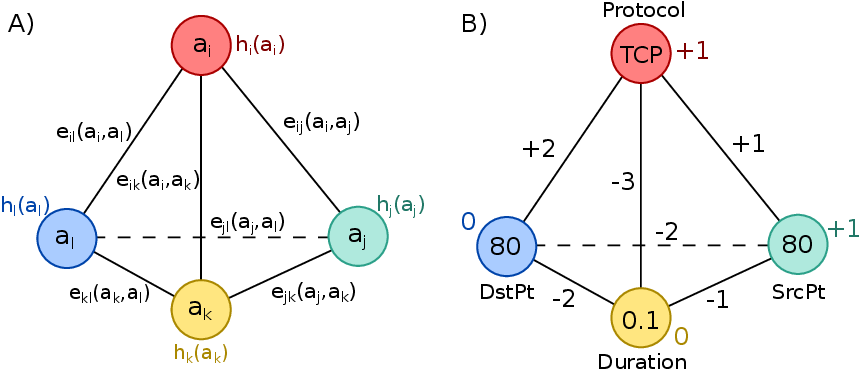}
\caption{A) Interacting spins on a crystalline lattice. B) Network flow mapped into a graph structure.}
\label{fig:model}
\vspace{0pt}
\end{figure}

EFC is based on inverse statistics. The main task of inverse statistics is to infer a statistical distribution based on a sample of it \cite{Cocco2018}.
Methods using inverse statistics have been successfully applied to problems in other disciplines, \eg predicting protein contacts in Biophysics \cite{Cocco2018, Morcos2011}.
Here, the statistical inference is based on the Potts model \cite{Wu1982}.
This model provides a mathematical description of interacting spins on a crystalline lattice. 
Within the model framework, interacting spins are mapped into a graph $\func$ (see Figure \ref{fig:model} A)), where each node $\nodei \in \node = \{1, ..., N\}$ has an associated spin $\spin_\nodei$, which can assume one value from a set $\alphabet$ that contains all possible individual quantum states.
Each node $\nodei$ has also an associated local field $\field_\nodei (\spin_\nodei)$ that is a function of $\spin_\nodei$'s state. 
Meanwhile, each edge $(\nodei,\nodej) \in \edge$, $\nodei, \nodej \in \node$, has an associated coupling value $\coup_{\nodei\nodej} (\spin_\nodei,\spin_{\nodej})$ that is a function of the states of spins $\spin_\nodei$ and $\spin_\nodej$ associated to nodes $\nodei$ and $\nodej$.
A specific system configuration has an associated total energy, determined by the Hamiltonian function $\mathcal{H}(a_1...a_N)$, which depends on all spin states.

In this work, we reuse the intuitions from the Potts model to characterize network flows (see Figure \ref{fig:model} B)).
An individual flow $\flowid$ is represented by a specific graph configuration $\funck$. Instead of spins, each node represents a selected feature $\featurei \in \node = \{Src Port, ..., Flags\}$. Within a given flow $\flowid$, each feature $\featurei$ assumes one value $\feature_{\flowid\featurei}$ from the set $\featureset_{\featurei}$ that contains all possible values for this feature. As in the Potts Model, each feature $\featurei$ has an associated local field $\field_\featurei(\feature_{\flowid\featurei})$. Meanwhile, $\edge = \{(\featurei,\featurej) | \featurei, \featurej \in \node ;  \featurei \neq \featurej\}$ is the set of edges determined by all possible pairs of features, creating a fully meshed graph that can represent different flow samples through their common features. Each edge has an associated coupling value determined by the function $\coup_{\featurei\featurej}(\feature_{\flowid\featurei}, \feature_{\flowid\featurej})$.

Since the values of local fields and couplings depend on the values assumed by features within a given flow, each distinct flow will have a different combination of these quantities. As in the Potts Model, the Hamiltonian involving local fields and couplings determines the total "energy" $\mathcal{H}(\feature_{\flowid1}...\feature_{\flowid \featureids})$ of each flow.
For instance, in Figure \ref{fig:model} B), the total "energy" of the flow is obtained by summing up all values associated with the edges and to the nodes, resulting in a total of -3.
Note that what we call energy is analogous to the notion of Hamiltonian in Quantum Mechanics.
It is important to note that the model described here is discrete, therefore continuous features must be discretized.
The classes for continuous feature discretization are shown in Section~\ref{sec:results}.
In the following, we present the framework applied to perform the statistical model inference and subsequent energy-based flow classification.

\subsection{Model inference} \label{sec:3a}
In this section, a statistical model is going to be inferred in terms of couplings and local field values to perform energy-based flow classification. The main idea consists in extracting a statistical model from benign flow samples to infer coupling and local field values that characterize this type of traffic. When calculating the energies of unlabeled flows using the inferred values, it is expected that benign flows will have lower energies than malicious flows.

Let $( A_{1}...A_{\featureids})$ be an $\featureids$-tuple of features, which can be instantiated for flow $\flowid$ as $(\feature_{\flowid 1}...\feature_{\flowid\featureids})$, with $\feature_{\flowid 1} \in \alphabet_1, ..., \feature_{\flowid\featureids} \in \alphabet_\featureids$.
Each feature value $\feature_{\flowid\featurei}$ is encoded by an integer from the set $\alphabet = \{ 1, 2, ..., Q \}$, \ie all feature alphabets are the same $\alphabet_\featurei = \alphabet$ of size $Q$. If a given feature can only assume $M$ values and $M < Q$, it is considered that values $M+1, ..., Q$ are possible, but will never be observed empirically. For instance, if the only possible values for feature \textit{protocol} are \{\textit{'TCP'}, \textit{'UDP'}\}, and given $Q = 4$. In this case, we would have the mapping \{\textit{'TCP'}:1, \textit{'UDP'}:2, \textit{' '}:3, \textit{' '}:4 \} and feature values 3 and 4 would never occur.  

Now, let $\flowidset$ be the set of all possible flows, \ie all possible combinations of feature values ($\flowidset = \alphabet^N$), and let $\flowset \subset \flowidset$ be a sample of flows. We can use inverse statistical physics to infer a statistical model associating a probability $P(\feature_{\flowid 1}...\feature_{\flowid \featureids})$ to each flow $\flowid \in \flowidset$ based on sample $\flowset$.
The global statistical model $P$ is inferred following the Entropy Maximization Principle \cite{Jaynes1957}:

\begin{align}
    \max_{P} & \ \ \ - \sum_{\flowid \in \flowidset}  \ \ \ \ \ \ \  P(\feature_{\flowid1}...\feature_{\flowid \featureids}) log (P(\feature_{\flowid 1}...\feature_{\flowid \featureids})) \label{eq:max}\\
    s.t.\nonumber\\
        & \ \ \ \sum_{\flowid \in \flowidset | \feature_{\flowid \featurei} = \feature_{\featurei}}  \ \ \ \ P(\feature_{\flowid 1} ... \feature_{\flowid \featureids}) = f_\featurei(\feature_\featurei) \label{eq:3} \\ 
        &\qquad\qquad\forall \featurei \in \node;\ \forall \feature_{\featurei} \in \Omega; \nonumber \\
        &\sum_{\flowid \in \flowidset | \feature_{\flowid \featurei} = \feature_{\featurei}, \feature_{\flowid \featurej} = \feature_{\featurej}} P(\feature_{\flowid 1} ... \feature_{\flowid \featureids}) = f_{\featurei \featurej} (\feature_\featurei, \feature_{\featurej}) \label{eq:4} \\
        &\qquad\qquad\forall (\featurei,\featurej) \in \node^2\ |\ \featurei \neq \featurej;\ \forall (\feature_{\featurei}, \feature_{\featurej}) \in \Omega^2;\nonumber
\end{align}
where $f_\featurei(\feature_\featurei)$ is the empirical frequency of value $\feature_\featurei$ on feature $\featurei$ and $f_{\featurei\featurej}(\feature_\featurei,\feature_{\featurej})$ is the empirical joint frequency of the pair of values $(\feature_\featurei,\feature_{\featurej})$ of features $\featurei$ and $\featurej$.
Note that constraints \ref{eq:3} and \ref{eq:4} force model $P$ to generate single as well as joint empirical frequency counts as marginals. This way, the model is sure to be coherent with empirical data.

Single and joint empirical frequencies $f_\featurei(\feature_\featurei)$ and $f_{\featurei\featurej}(\feature_\featurei,\feature_{\featurej})$ are obtained from set $\flowset$ by counting occurrences of a given feature value $\feature_\featurei$ or feature value pair $(\feature_\featurei, \feature_\featurej)$, respectively, and dividing by the total number of flows in $\flowset$.
Since the set $\flowset$ is finite and much smaller than $\flowidset$, inferences based on $\flowset$ are subjected to undersampling effects.
Following the theoretical framework proposed in \cite{Morcos2011}, we add pseudocounts to empirical frequencies to limit undersampling effects by performing the following operations:
    \begin{equation}
        f_\featurei(\feature_\featurei) \leftarrow (1 - \alpha ) f_\featurei(\feature_\featurei) + \frac{\alpha}{Q}\label{eq:single_site}
    \end{equation}
    \begin{equation}
        f_{\featurei \featurej}(\feature_\featurei,\feature_{\featurej}) \leftarrow (1 - \alpha ) f_{\featurei \featurej}(\feature_\featurei,\feature_{\featurej}) + \frac{\alpha}{Q^2}\label{eq:pair_freq}
    \end{equation}
where $(\feature_\featurei, \feature_{\featurej}) \in \alphabet^2$ and $0 \leq \alpha \leq 1$ is a parameter defining the weight of the pseudocounts.
The introduction of pseudocounts is equivalent to assuming that $\flowset$ is extended with a fraction of flows with uniformly sampled features.

The proposed maximization can be solved using a Lagrangian function such as presented in \cite{Jaynes1957}, yielding the following Boltzmann-like distribution: 
    \begin{equation}
        P^*(\feature_{\flowid1}...\feature_{\flowid\featureids}) = \frac{e^{- \mathcal{H}(\feature_{\flowid1}...\feature_{\flowid\featureids})}}{Z} \label{eq:opt_dist}
    \end{equation}
where
    \begin{equation}
        \mathcal{H}(\feature_{\flowid1}...\feature_{\flowid\featureids}) = - \sum_{\featurei,\featurej \mid \featurei<\featurej} \coup_{\featurei\featurej}(\feature_{\flowid\featurei},\feature_{\flowid\featurej}) - \sum_\featurei \field_\featurei(\feature_{\flowid\featurei}) \label{eq:hamil}
    \end{equation}
is the Hamiltonian of flow $\flowid$ and $Z$ (eq. (\ref{eq:opt_dist})) is the partition function that normalizes the distribution. Since in this work we are not interested in obtaining individual flow probabilities, $Z$ is not required and, as a consequence, its calculation is omitted. Our objective is to calculate individual flows energies, \ie individual Hamiltonians as determined in eq. (\ref{eq:hamil}).

Note that the Hamiltonian, as presented above, is fully determined regarding the Lagrange multipliers $ \coup_{\featurei \featurej}(\cdot)$ and $\field_\featurei(\cdot)$ associated to constraints (\ref{eq:3}) and (\ref{eq:4}), respectively. Within the Potts Model framework, the Lagrange multipliers have a special meaning, with the set $\{\coup_{\featurei \featurej}(\feature_\featurei,\feature_{\featurej}) | (\feature_\featurei, \feature_\featurej) \in \alphabet^2\}$ being the set of all possible coupling values between features $i$ and $j$ and $\{\field_\featurei(\feature_\featurei) | \feature_\featurei \in \alphabet\}$ the set of possible local fields associated to feature $\featurei$.

Inferring the local fields and pairwise couplings is difficult since the number of parameters exceeds the number of independent constraints.
Due to the physical properties of interacting spins, it is possible to infer pairwise coupling values $\coup_{\featurei \featurej}(\feature_\featurei,\feature_{\featurej})$ using a Gaussian approximation.
Assuming that the same properties apply for flow features, we infer coupling values as follows:
    \begin{align}
        \coup_{\featurei \featurej}(\feature_\featurei,\feature_{\featurej}) = -(C^{-1})_{\featurei \featurej} (\feature_\featurei,\feature_{\featurej}),\label{eq:coupling}\\ \forall (\featurei, \featurej) \in \node^2, \forall (\feature_\featurei, \feature_\featurej) \in \alphabet^2, \feature_\featurei, \feature_\featurej \neq Q \nonumber
    \end{align}
where
    \begin{equation}
        C_{\featurei \featurej}(\feature_\featurei,\feature_{\featurej}) = f_{\featurei\featurej}(\feature_\featurei,\feature_{\featurej}) - f_\featurei(\feature_\featurei)f_{\featurej}(\feature_{\featurej})
    \end{equation}
is the covariance matrix obtained from single and joint empirical frequencies. Taking the inverse of the covariance matrix is a well known procedure in statistics to remove the effect of indirect correlation in data \cite{giraud1999superadditive}. Now, it is important to clarify that the number of independent constraints in eq. (\ref{eq:3}) and eq. (\ref{eq:4}) is actually  $\frac{N(N-1)}{2}(Q-1)^2 + N(Q-1)$, even though the model in eq. (\ref{eq:opt_dist}) has $\frac{N(N-1)}{2}Q^2 + NQ$ parameters. So, without loss of generality, we set:
\begin{align}
    \coup_{\featurei, \featurej}(\feature_\featurei, Q) = \coup_{\featurei, \featurej}(Q, \feature_\featurej) = \field_\featurei (Q) = 0 
\end{align}
Thus, in eq. (\ref{eq:coupling}) there is no need to calculate $\coup_{\featurei, \featurej}(\feature_\featurei, \feature_\featurej)$in case $\feature_\featurei$ or $\feature_\featurej$ is equal to $Q$ \cite{Morcos2011}.
Afterwards, local fields $\field_\featurei(\feature_\featurei)$ can be inferred using a mean-field approximation \cite{georges1991expand}:
    \begin{align}
        \frac{f_\featurei(\feature_\featurei)}{f_\featurei(Q)} = exp \left ( \field_\featurei(\feature_\featurei) + \sum_{\featurej,\feature_{\featurej}} \coup_{\featurei \featurej}(\feature_\featurei,\feature_{\featurej}) f_{\featurej}(\feature_{\featurej}) \right ), \label{eq:mf}\\  \forall \featurei \in \node, \feature_\featurei \in \alphabet, \feature_\featurei \neq Q \nonumber
    \end{align}
where $f_\featurei(Q)$ is the frequency of the last element $\feature_\featurei = Q$ for any feature $\featurei$ used for normalization. It is also worth mentioning that the element Q is arbitrarily selected and could be replaced by any other value in \{1$\dots$Q\} as long as the selected element is kept the same for calculations of the local fields of every feature $\featurei \in \node$.
Note that in eq. (\ref{eq:mf}) the empirical single frequencies $f_\featurei(\feature_\featurei)$ and the coupling values $\coup_{\featurei\featurej}(\feature_\featurei,\feature_{\featurej})$ are known, yielding:
    \begin{align}
        \field_\featurei(\feature_\featurei) = ln \left (\frac{f_\featurei(\feature_\featurei)}{f_\featurei(Q)} \right ) - \sum_{\featurej, \feature_\featurej}\coup_{\featurei\featurej}(\feature_\featurei, \feature_\featurej)f_\featurej(\feature_\featurej) \label{eq:mf2}
    \end{align}
In the mean-field approximation presented above, the interaction of a feature with its neighbors is replaced by an approximate interaction with an averaged feature, yielding an approximated value for the local field associated to it. 

For further details about these calculations, please refer to \cite{Cocco2018}.
Now that all model parameters are known, it is possible to calculate a given flow energy according to eq. (\ref{eq:hamil}). In the following, we are going to present the theoretical framework implementation to perform a two-class, \ie benign and malicious, flow classification, \ie Energy-based flow classification (EFC).

\subsection{Energy-based flow classification} \label{sec:3b}
The energy of a given flow can be calculated according to eq.~(\ref{eq:hamil}) based on the values of its features and the parameters from the statistical model inferred in Section~\ref{sec:3a}.
A given flow energy is the negative sum of couplings and local fields associated with its features, according to a given statistical model.
It means that a flow that resembles the ones used to infer the model is likely to be low in energy.

Since \ac{EFC} is an anomaly-based classifier, the statistical model used for classification is inferred based only on benign flow samples.
We would then expect the energies of benign samples to be lower than the energies of malicious samples.
In other words,  what the energy value of a given flow is capturing is how dissimilar that flow is to a set of known benign flows used to infer the model in the training phase.
In terms of frequencies, this means that, if a given flow presents feature values combinations that are very frequent in benign flow samples, its energy will be low.
In this sense, it is possible to classify flow samples as benign or malicious based on a chosen energy threshold.
The classification is performed by stating that samples with energy smaller than the threshold are benign, and samples with energy greater than or equal to the threshold are malicious.
Note that the threshold for classification can be chosen in different ways, and it can be static or dynamic.
In this work, we will consider a static threshold.

Algorithm \ref{algo:efc} shows the implementation of \ac{EFC}. In lines 2-5, the statistical model for the sampled flows is inferred, as described by eqs.~(\ref{eq:single_site}), (\ref{eq:pair_freq}), (\ref{eq:coupling}) and (\ref{eq:mf2}).
Afterward, on lines 6-27, the classifier monitors the network waiting for a captured flow. When a flow is captured, its energy is calculated on lines 9-20, according to the Hamiltonian in eq.~(\ref{eq:hamil}).
The computed flow energy is compared to a known threshold (\textit{cutoff}) value on line 21.
In case the energy falls above the threshold, the flow is classified as malicious and should be forwarded to deep packet inspection (line 23) for assessment.
Otherwise, the flow is released, and the classifier waits for another flow.

\begin{algorithm}[!h]
    \hspace*{\algorithmicindent}
    \textbf{Input:} $benign\_flows_{(K \times N)}$, $Q$, $\alpha$, $cutoff$
    
    \begin{algorithmic}[1]
        \State import all model inference functions
        \State $f\_i \leftarrow SiteFreq(benign\_flows, Q, \alpha)$
        \State $f\_ij \leftarrow PairFreq(benign\_flows, f\_i, Q, \alpha)$
        \State $e\_ij \leftarrow Couplings(f\_i, f\_ij, Q)$
        \State $h\_i \leftarrow LocalFields(e\_ij, f\_i, Q)$
        
        \While{Scanning the Network}
            \State $flow \leftarrow$ wait\_for\_incoming\_flow()
            \State e $\leftarrow 0$
            \For{$i$ $\leftarrow 1$ to $N-1$}
                \State $a\_i \leftarrow flow[i]$
                \For{$j$ $\leftarrow i + 1$ to $N$}
                    \State $a\_j \leftarrow flow[j]$
                    \If{$a\_i \neq Q$ and $a\_j \neq Q$}
                        \State $e \leftarrow e - e\_ij[i,a\_i,j,a\_j]$
                    \EndIf
                \EndFor
                \If{$a\_i \neq Q$}
                    \State $e \leftarrow e - h\_i[i,a\_i]$
                \EndIf
            \EndFor
            \If{$e \geq cutoff$}
                \State stop\_flow()
                \State forward\_to\_DPI()
            \Else
                \State release\_flow()
            \EndIf
        \EndWhile
    \end{algorithmic}\vspace{-2pt}
    \caption{Energy-based Flow Classifier}
    \label{algo:efc}
\end{algorithm}

It is essential to highlight that the time complexity of the training step of \ac{EFC} is $O((M \times Q)^3 + N \times M^2 \times Q^2)$, where $N$ is the number of samples, $M$ is the number of features, and $Q$ is the size of the alphabet.
Meanwhile, the complexity of the classification step for each sample is $O(M^2)$.
It means that, in both steps, the complexity is more dependant on the number of features chosen, which can be kept small by using a feature selection mechanism, \eg \ac{PCA}.
However, we do not currently explore any feature selection mechanisms because we consider it to be out of scope of this work, in which the main aim is only to present a first version of our newly proposed classifier for NIDS.

\begin{figure}[H]
\centering
\includegraphics[width=\columnwidth, clip]{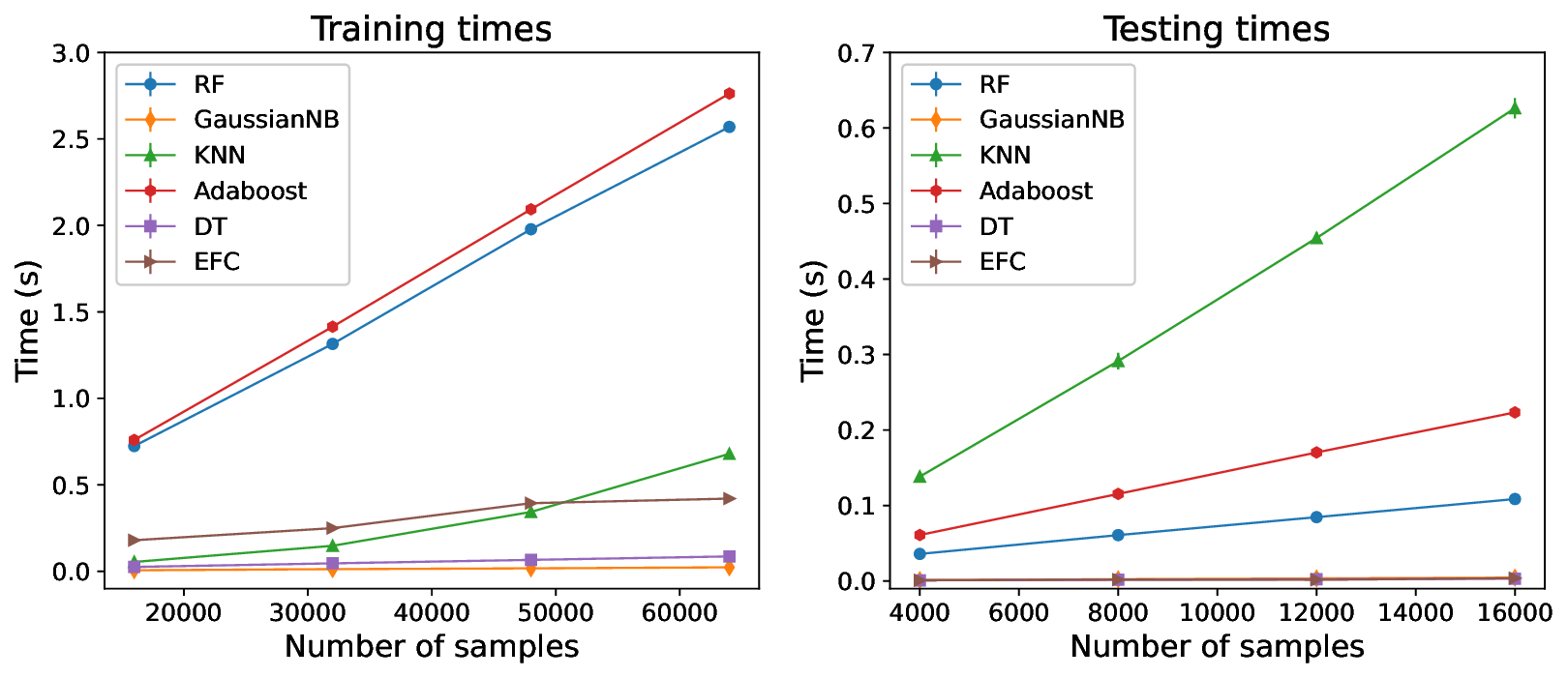}
\caption{Training and testing runtimes of different ML-based classifiers (scikit-learn implementation) and EFC (Cython implementation) over the CIDDS-001 dataset.}
\label{fig:times}
\vspace{0pt}
\end{figure}

\ac{EFC} has a low training cost, linear in the number of samples ($N$), when compared to some ML-based classifiers, such as Adaboost (AB), Random Forest (RF), \ac{SVM}, and \ac{MLP} \cite{buczak2015survey}.
EFC's training complexity is considered to be dominated by the term $NM^2Q^2$ because the number of training samples is expected to be much bigger than both the number of features $M$ and the size of the alphabet $Q$, which means that the term $(MQ)^3$ is likely not dominant over $NM^2Q^2$.
Figure~\ref{fig:times} shows a comparative analysis of training and testing runtimes of different ML-based classifiers and EFC performed on the CIDDS-001 dataset.
A Cython implementation of EFC\footnote{EFC - \url{https://cyberseclab.gigacandanga.net.br/CyberSecLab/one-class-efc}} was compared to the scikit-learn\footnote{Scikit-learn library - \url{https://scikit-learn.org}} implementation of the other classifiers in a Ubuntu 20.04 OS using a single thread.
The MLP and the SVM classifiers were omitted from the plot for visualization purposes, since their training time complexity is not linear on the number of samples.
The results obtained are consistent with the analytical complexity analysis.

Considering the implementation shown in this section, in the following we present the results obtained using \ac{EFC} and ML-algorithms in classification experiments.


\section{Results} \label{sec:results}
In this section, we present the results obtained for \ac{EFC} and ML-based classifiers in different binary classification experiments considering three different datasets, \ie CIDDS-001, CICIDS17, and CICDDoS19.
First, we show that EFC can separate benign from malicious flows based on their energies, a result that is consistent for all considered datasets.
Then, we present EFC's classification performance and compare it to the classification performance of ML-based classifiers in different experiments.

It is important to highlight that the classification experiments we perform in this work were designed not only to assess the performance of different classifiers but also to investigate their capability of adaptation to different domains, \ie data distributions.
Hence, we performed two kinds of experiments: training/testing in the same domain, and training/testing in different domains.
For training/testing in the same domain, in each experiment, we assessed the average performance of the classifiers over ten different test sets, containing 10,000 benign and 10,000 malicious samples each, randomly selected from the full dataset.
Models were inferred based on 80\% the test set and tested on the remaining 20\%.
The inferred models were used for each experiment to assess the performance of the classifiers over another ten test sets composed of 2,000 benign and 2,000 malicious samples from another domain (data distribution).

\begin{figure*}[ht]
    \centering
    \includegraphics[width=\textwidth, clip]{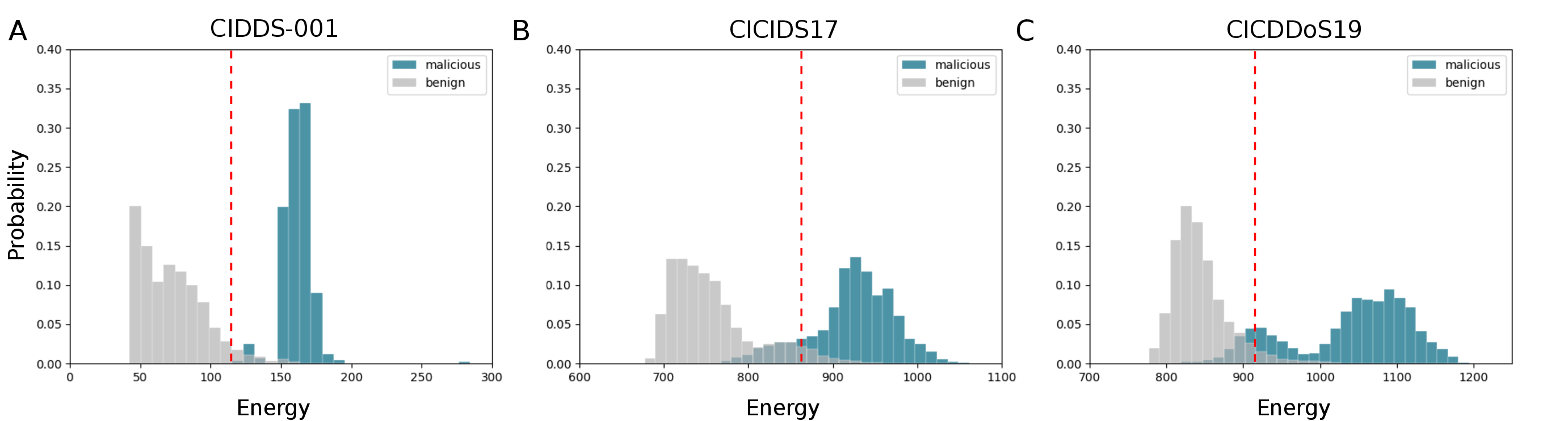}
    \caption{Energy histograms of benign (n = 20,000 in each plot) and malicious (n = 20,000 in each plot) flow samples obtained in the testing phase of a classification experiment performed over the CIDDS-001 dataset (A), CICIDS17 (B) and CICDDoS19 (C) datasets. The energy threshold for classification is shown as a red dashed line and corresponds to the 95th percentile of the energy distribution obtained in the training phase.}
    \label{fig:energies}
    \vspace{-12pt}
\end{figure*}

\subsection{EFC characterization}
To assess EFC capabilities to separate benign from malicious traffic flow samples correctly, we performed classification experiments considering datasets CIDDS-001, CICIDS17 and CICDDoS19.
First, we inferred models based on benign samples from the OpenStack (simulated) environment within the CIDDS-001 dataset.
This models were used to calculate the energy of different benign and malicious flow samples coming also from the simulated traffic.
Figure~\ref{fig:energies}A shows energy values of 40,000 classified flow samples, a merge of the results obtained over ten randomly sampled test sets, as described in the last paragraph of the previous section.
The statistical model used to calculate the energies in each test set  was inferred based on 8,000 benign flows randomly sampled from the simulated traffic.
Flow samples with energy values falling above the energy cutoff, defined as the 95th percentile of the benign traffic training distribution (red dashed line), would be classified as malicious, while the remaining samples would be classified as benign.
It is possible to observe that the separation between the two flow classes is clear, \ie the energy distribution of tested benign flows falls mostly on the left side of the cutoff line, while the energy distribution of tested malicious flows falls mostly on the right side of the cutoff line, as expected.

Figure~\ref{fig:energies}B-C shows the results of analogous experiments performed on the remaining two datasets.
Similarly to what is observed for CIDDS-001, when trained on CICIDS17, EFC is also capable of clearly separating the two classes (Figure~\ref{fig:energies}B).
This means that the benign energy histogram of tested samples falls mostly on the left side of the cutoff line (95th percentile of the training distribution), while the malicious energy histogram of tested samples falls mainly on the right side of the cutoff line.
Again, the same result can be observed for dataset CICDDoS19 (Figure~\ref{fig:energies}C).
It is interesting to observe that, although the benign flows energy histograms looks similar regarding variance for the three datasets, the malicious flows energy histogram varies.
In CIDDS-001, it has very low variance, reflecting the fact that this dataset contain only four classes of attacks and is highly imbalanced, while in CICIDS17 and CICDDoS19, malicious energy histograms have a broader spread, reflecting the greater variability of malicious flows that exists in those datasets.

\begin{figure}[h!]
\centering
\includegraphics[width=\columnwidth, clip]{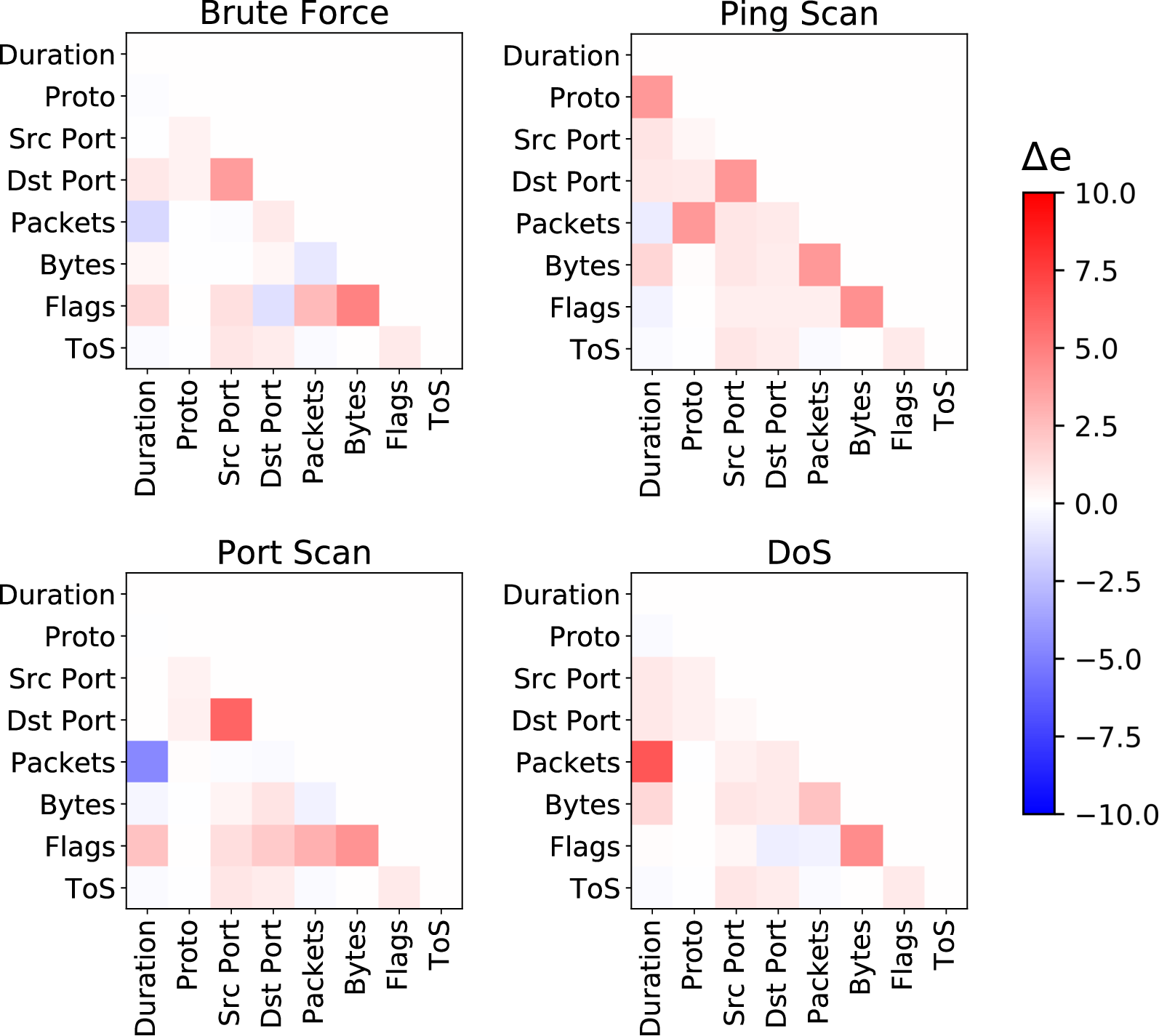}
\caption{Individual contributions of each feature pair for the total energy of each attack type within CIDDS-001 dataset: brute force (n = 150), ping scan (n = 100), port scan (n = 800) and DoS (n = 8,950). The heatmap shows the energetic difference $\Delta e$ that each pair of features has in relation to the average expected energy value of that pair in benign flows (n = 10,000), calculated as $\Delta e = \bar e_{attack} - \bar e_{benign}$.}
\label{fig:couplings}
\vspace{0pt}
\end{figure}

The white box nature of the statistical model inferred by EFC is demonstrated in Figure~\ref{fig:couplings}, where the energy of different attack classes is broken down to the individual contributions of each pair of features.
As shown, for a given attack class, it is possible to identify which combination of features is contributing the most to its abnormality (red squares) and which pairs of features are similar to normal traffic (blue squares) and might be confounding the model.
This analysis was done considering only the couplings and not the local fields. 
It is interesting to note that different kinds of attacks are characterized by different combinations of abnormal feature pairs, as expected.
For instance, the most abnormal thing about DoS attacks is the combination of number of packets and duration, which is consistent with spoofed source address flooding attacks.
In turn, Port Scan attacks present an abnormal combination of source and destination ports, in contrast with packets and duration, which are exceedingly normal.
Ping Scan attacks are characterized by the protocol ICMP coupled with smaller than usual duration and number of packets, an abnormal combination of source and destination ports, and the absence of flags coupled with a small number of bytes.
As for Brute Force, the abnormal coupling between source and destination ports is consistent with unvarying destination and evolving source port, along with abnormally coupled flags and bytes, which reflect short TCP connections.

An analogous breakdown can be done for individual flow samples, allowing for the understanding of which features cause a specific sample to be classified as malicious or benign.
This means that the statistical model inferred by EFC allows a direct mapping between specific features of a given flow (model input), \eg ICMP protocol, and a single packet, and the increase in energy they represent (model output), providing an understanding of \textit{why} a given flow was classified as malicious or benign.
Assume, for instance, that EFC classified a flow as an anomaly and the network manager wants to understand the reason for that classification.
This is equivalent to understanding \textit{why} that specific flow generated that specific energy value.
As the energy of a flow is the sum of the couplings and local fields (functions of the flow features), it is possible to verify which factors of that sum contributed the most to its increase. Consequently, the manager should be able to understand which features contributed the most to classifying the flow as an attack. Now, observing the value assumed by these features in that specific flow, it is possible to clearly understand the classification: some of those features are unlikely to appear in benign flows (considering the normal behavior inferred from the training set).

In summary, the results presented in this subsection show that EFC can correctly discriminate between the two flow classes considered, \ie benign and malicious, consistently across all datasets considered.
In addition, it was shown how the total energy of different attack classes can be broken down and analyzed in detail.
This is illustrative of the white box nature of the statistical model inferred by EFC.
In the following, classification results are shown for different classifiers and compared with the results obtained for \ac{EFC}.

\subsection{Comparative analysis of EFC's performance}
We compared EFC to five different \ac{ML} classifiers: K-Nearest Neighbors (KNN) \cite{hartigan1979algorithm}, Decision Tree (DT) \cite{quinlan1987simplifying, swain1977decision},
Multilayer perceptron (MLP) \cite{mcculloch1943logical}, Naive Bayes (NB) \cite{lewis1998naive}, and Support Vector Machine (SVM) \cite{cortes1995support}, all deployed with their default scikit-learn configurations.
Additionally, two ensemble methods, namely Adaboost (AB) \cite{freund1996experiments} and Random Forest (RF) \cite{svetnik2003random}, were considered, also with default scikit-learn parameters.
Flow features were only discretized for EFC (Table~\ref{tab:classes}) since discretization would impair the performance of most \ac{ML} algorithms.
The metrics used to compare the results were the F1 score and the area under the ROC curve (AUC).
The first metric, F1 score, is the harmonic mean of the Precision and the Recall, \ie
\begin{align}
    F1 = \frac{2}{Precision^{-1} + Recall^{-1}} = \frac{2 \cdot Precision \cdot Recall}{Precision + Recall}
\end{align}
where $Precision = TP / (TP+FP)$, $Recall = TP / (TP+FN)$, TP are the true positives, \ie malicious traffic classified as malicious, FP are the false positives, \ie benign traffic classified as malicious, and FN are the false negatives, \ie malicious traffic classified as benign.
The second metric, the area under the ROC curve (AUC), is one of the most widespread evaluation metrics for binary classifiers \cite{japkowicz2011evaluating, brzezinski2017prequential}.
The ROC curve is constructed by plotting the true positive rate (TPR) against the false positive rate (FPR) at different classification thresholds.
It means that the AUC is the probability that a randomly chosen positive example will receive a higher score than a randomly chosen negative one.
One of the main advantages of the AUC is that it is invariant to changes in class distribution, \ie the ROC curve will not change if the distribution changes in a test set, but the underlying conditional distributions from which the data are drawn stay the same \cite{wu2007improved, brzezinski2017prequential}.
Since we are interested in evaluating domain adaptation, this metric is particularly interesting to be adopted in this work.

Table \ref{tab:classes} shows the classes considered for feature discretization on CIDDS-001 dataset.
Since \textit{TCP Flags} is the discrete feature with more possible values (32 possibilities), the alphabet size $Q$ was set to 32.
Each continuous feature values were clustered in a certain number of classes (or bins), up to $Q$ classes.
Classes were determined in such a way that the number of values within each class was similar for all classes.
Features within datasets CICIDS17 and CICDDoS19 were also discretized in such a way that the number of values within each bin was similar for all bins.
These discretizations are not shown here because of the high number of features (~80 features) in these datasets.

\begin{table}[!h]
\renewcommand{\arraystretch}{1.3}
\caption{Classes considered for feature discretization on CIDDS-001}
\label{tab:classes}
\centering
\begin{adjustbox}{max width=\columnwidth}
\begin{tabular}{ l  c }
\hline
Feature & List of classes upper limits \\
\hline
Duration &  0.001, 0.002, 0.003, 0.004, 0.005, 0.006, 0.01, 0.04, 1, 10, 100, $\infty$\\
Protocol & TCP, UDP, GRE, ICMP, IGMP \\
Src Port & 50, 60, 100, 400, 500, 40000, 60000, $\infty$ \\
Dst Port & 50, 60, 100, 400, 500, 40000, 60000, $\infty$  \\
Num. of Bytes &  50, 60, 70, 90, 100, 110, 200, 300, 400, 500, 700, 1000, 5000, $\infty$ \\
Num. of Packets &  2, 3, 4, 5, 6, 7, 10, 20, $\infty$ \\
TCP Flags &  $\{(f_1, f_2, f_3, f_4, f_5) | f_i \in \{0,1\}\}$ \\
\hline
\end{tabular}
\end{adjustbox}
\end{table}

To evaluate EFC's performance compared to other classifiers, we performed three independent experiments.
The first experiment was performed on CIDDS-001.
Training was performed on simulated flow samples, while testing was performed on both simulated and real flow samples captured in an external server.
The second and the third experiments were cross-dataset experiments performed on CICIDS17 and CICDDoS19.
In the former, training was performed on CICIDS17, with testing on both datasets, while in the latter training was performed on CICDDoS19, with testing on both datasets.

Essentially, in each experiment, we measured the performance of the classifiers when trained and tested in the same domain and when trained in one domain and tested in a different one.
The performance was measured as the average over ten different test sets, composed of 10,000 benign and 10,000 malicious samples each, randomly selected form the full dataset.
With 80\% of each test set being used for training and 20\% for testing.
The test sets containing samples from a different domain were not used for training, hence they were composed of only 2,000 benign and 2,000 malicious samples, randomly selected from the full dataset.
EFC's cutoff was defined to be at the 95th percentile of the energy distribution obtained in the training phase based solely in benign samples. It means that we used a completely statistical threshold based only in the training considering benign traffic without adjustments based on malicious samples, such as other ML-Algorithms require.
The average composition of the test sets is shown in Table~\ref{tab:sets001} and \ref{tab:sets1719}.

\begin{table}[!h]
\renewcommand{\arraystretch}{1.3}
\caption{Average composition of each of the test sets in experiment 1}
\label{tab:sets001}
\centering
\begin{adjustbox}{max width=\columnwidth}
\begin{tabular}{ l  r c  l  r }
\hline
\multicolumn{2}{ c }{CIDDS-001 OpenStack} & & \multicolumn{2}{ c }{CIDDS-001 real traffic} \\
\cline{1-2}
\cline{4-5}
Label & Number & & Label & Number\\
\hline
\textit{normal} & 10,000 & & \textit{unknown} & 2,000 \\
\textit{dos} & 9,800 & & \textit{suspicious} & 2,000 \\
\textit{pingScan} & 20 & & & \\
\textit{portScan} & 150 & & & \\
\textit{bruteForce} & 30 & & & \\
\textbf{Total} & 20,000 & & \textbf{Total} & 4,000 \\
\hline
\end{tabular}
\end{adjustbox}
\end{table}

\begin{table}[!h]
\renewcommand{\arraystretch}{1.3}
\caption{Average composition of each of the test sets in cross-dataset experiments 2 and 3}
\label{tab:sets1719}
\centering
\begin{adjustbox}{max width=\columnwidth}
\begin{tabular}{ l  r c  l  r }
\hline
\multicolumn{2}{ c }{CICIDS17} & & \multicolumn{2}{ c }{CICDDoS19} \\
\cline{1-2}
\cline{4-5}
Label & Number & & Label & Number\\
\hline
\textit{benign} & 10,000 & & \textit{benign} & 10,000\\
\textit{FTP Patator} & 170 & & \textit{DrDoS DNS} & 890 \\
\textit{SSH Patator} & 80 & & \textit{DrDoS LDAP} & 370  \\
\textit{DDoS} & 2,740 & & \textit{DrDoS MSSQL} & 890  \\
\textit{PortScan} & 1,060 & & \textit{DrDoS NetBIOS} & 880  \\
\textit{Bot} & 110 & & \textit{DrDoS NTP} & 890 \\
\textit{Infiltration} & 20 & & \textit{DrDoS SNMP} & 200  \\
\textit{Brute force} & 50 & & \textit{DrDoS SSDP} & 890 \\
\textit{SQL injection} & 10 & & \textit{DrDoS UDP} & 890 \\
\textit{XSS} & 10 & & \textit{Syn} & 890  \\
\textit{DoS Hulk} & 2,730 & & \textit{TFTP} & 890 \\
\textit{DoS GoldenEye} & 2,730 & & \textit{LDAP} & 120 \\
\textit{DoS Slowloris} & 120 & & \textit{NetBIOS} & 140  \\
\textit{DoS Slowhttptest} & 170 & & \textit{MSSQL} &  660 \\
& & & \textit{Portmap} & 400 \\
& & & \textit{UDP} & 880 \\
& & & \textit{UDPLag} & 120 \\
\textbf{Total} & 20,000 & & \textbf{Total} & 20,000  \\
\hline
\end{tabular}
\end{adjustbox}
\vspace{0pt}
\end{table}

Table~\ref{tab:avg_results} shows the average performance and standard error (95\% confidence interval) of each classifier on the first experiment considering CIDDS-001 dataset.
When trained and tested in the same simulated environment, DT is the algorithm presenting the best performance, with an F1-score of $0.999 \pm 0.000$ and $0.999 \pm 0.000$ AUC.
EFC does also perform well, being the second best in terms of AUC ($0.997 \pm 0.001$).
When trained in the simulated environment and tested in the real environment, EFC outperforms the other classifiers (both simple and ensemble methods) in F1-score ($0.675 \pm 0.009$) and AUC ($0.720 \pm 0.001$).
It is noteworthy that all algorithms present a considerable degradation in performance when tested in a different domain, showing how sensitive the inferred models are to changes in data distribution.

\begin{table}[!h]
\renewcommand{\arraystretch}{1.3}
\caption{Average classification performance and standard error (95\%~CI) - training performed on CIDDS-001 simulated traffic}
\label{tab:avg_results}
\centering
\begin{adjustbox}{max width=\columnwidth}
\begin{tabular}{ l c c c c}
\hline
 & \multicolumn{2}{c }{Train/Test simulated} & \multicolumn{2}{c }{Train simulated/Test real} \\
\cline{2-3}
\cline{4-5}
Classifier & F1 score & AUC & F1 score & AUC \\
\hline
NB & 0.043 $\pm$ 0.016 & 0.502 $\pm$ 0.004 & 0.057 $\pm$ 0.024 & 0.517 $\pm$ 0.016 \\ 
KNN & 0.988 $\pm$ 0.001 & 0.994 $\pm$ 0.000 & 0.118 $\pm$ 0.014 & 0.524 $\pm$ 0.001 \\ 
DT & \textbf{0.999 $\pm$ 0.000} & \textbf{0.999 $\pm$ 0.000} & 0.556 $\pm$ 0.007 & 0.619 $\pm$ 0.000 \\ 
SVM & 0.805 $\pm$ 0.003 & 0.951 $\pm$ 0.002 & 0.531 $\pm$ 0.005 & 0.707 $\pm$ 0.003 \\ 
MLP & 0.979 $\pm$ 0.002 & 0.993 $\pm$ 0.001 & 0.151 $\pm$ 0.016 & 0.596 $\pm$ 0.002 \\ 
\textbf{EFC} & 0.975 $\pm$ 0.001 & 0.997 $\pm$ 0.001 & \textbf{0.675 $\pm$ 0.009} & \textbf{0.720 $\pm$ 0.001} \\
\hline
Ensemble \\
\hline
AB & 0.999 $\pm$ 0.000 & 1.000 $\pm$ 0.000 & 0.594 $\pm$ 0.022 & 0.630 $\pm$ 0.000 \\ 
RF & 0.999 $\pm$ 0.000 & 1.000 $\pm$ 0.000 & 0.269 $\pm$ 0.018 & 0.714 $\pm$ 0.000 \\ 

\hline
\end{tabular}
\end{adjustbox}
\end{table}

Table \ref{tab:compinter1} shows the results of experiment two, which was performed on CICIDS17 and CICDDoS19 datasets.
When trained and tested on CICIDS17, DT is again the algorithm to present the best performance both in terms of F1-score ($0.994 \pm 0.001$) and AUC ($0.994 \pm 0.001$), though indistinguishable from MLP AUC ($0.993 \pm 0.001$).
Notably, when trained on CICIDS17 and tested on CICDDoS19, EFC outperformed the other simple algorithms in both F1-score ($0.787 \pm 0.004$) and AUC ($0.781 \pm 0.003$).
Again, it is possible to see that EFC is the best algorithm in adapting to a different data distribution when evaluating both metrics.
However, when considering also ensemble methods, RF outperforms EFC, which becomes second best in terms of AUC.

\begin{table}[!h]
\renewcommand{\arraystretch}{1.3}
\caption{Average classification performance and standard error (95\%~CI) - training performed on CICIDS17}
\label{tab:compinter1}
\centering
\begin{adjustbox}{max width=\columnwidth}
\begin{tabular}{ l  c  c  c  c }
\hline
 & \multicolumn{2}{ c }{Train/Test CICIDS17} & \multicolumn{2}{ c }{Train CICIDS17/Test CICDDoS19}  \\
\cline{2-3}
\cline{4-5}
Classifier & F1 score & AUC & F1 score & AUC \\
\hline
NB & 0.344 $\pm$ 0.049 & 0.413 $\pm$ 0.021 & 0.344 $\pm$ 0.049 & 0.413 $\pm$ 0.049 \\ 
KNN & 0.961 $\pm$ 0.001 & 0.987 $\pm$ 0.001 & 0.457 $\pm$ 0.046 & 0.771 $\pm$ 0.001 \\ 
DT & \textbf{0.994 $\pm$ 0.001} & \textbf{0.994 $\pm$ 0.001} & 0.168 $\pm$ 0.090 & 0.525 $\pm$ 0.001 \\ 
SVM & 0.930 $\pm$ 0.003 & 0.974 $\pm$ 0.001 & 0.264 $\pm$ 0.025 & 0.664 $\pm$ 0.003 \\ 
MLP & 0.961 $\pm$ 0.003 & \textbf{0.993 $\pm$ 0.001} & 0.221 $\pm$ 0.033 & 0.775 $\pm$ 0.003 \\
\textbf{EFC} & 0.898 $\pm$ 0.003 & 0.975 $\pm$ 0.001 & \textbf{0.787 $\pm$ 0.004} & \textbf{0.781 $\pm$ 0.003} \\ 
\hline
Ensemble \\
\hline
AB & 0.991 $\pm$ 0.002 & 1.000 $\pm$ 0.000 & 0.228 $\pm$ 0.055 & 0.698 $\pm$ 0.002 \\ 
RF & 0.997 $\pm$ 0.001 & 1.000 $\pm$ 0.000 & 0.021 $\pm$ 0.003 & 0.867 $\pm$ 0.001 \\ 
\hline
\end{tabular}
\end{adjustbox}
\end{table}

Further, Table \ref{tab:compinter2} shows the results of experiment three, which was performed on CICIDS17 and CICDDoS19 datasets.
Once more, DT overperformed the other classifiers when training and testing on the same dataset, with F1-score of $0.998 \pm 0.000$ and AUC of $0.998 \pm 0.000$.
When tested on the CICDDoS19 dataset though, EFC achieved the best F1-score ($0.641 \pm 0.002$), while KNN was the best in terms of AUC ($0.670 \pm 0.002$).
EFC's AUC ($0.664 \pm 0.002$) was the second best, which means that EFC performance was good, taking into consideration both the F1-score and the AUC.
Even though this adaptation seems more challenging than the previous ones, EFC's performance was consistent on all the experiments performed.
 
\begin{table}[!h]
\renewcommand{\arraystretch}{1.3}
\caption{Average classification performance and standard error (95\%~CI) - training performed on CICDDoS19}
\label{tab:compinter2}
\centering
\begin{adjustbox}{max width=\columnwidth}
\begin{tabular}{ l c c c c}
\hline
 & \multicolumn{2}{c }{Train/Test CICDDoS19} & \multicolumn{2}{c }{Train CICDDoS19/Test CICIDS17} \\
\cline{2-3}
\cline{4-5}
Classifier & F1 score & AUC & F1 score & AUC \\
\hline
NB & 0.590 $\pm$ 0.006 & 0.428 $\pm$ 0.007 & 0.590 $\pm$ 0.006 & 0.428 $\pm$ 0.006 \\ 
KNN & 0.960 $\pm$ 0.002 & 0.984 $\pm$ 0.001 & 0.397 $\pm$ 0.043 & \textbf{0.670 $\pm$ 0.002} \\ 
DT & \textbf{0.998 $\pm$ 0.000} & \textbf{0.998 $\pm$ 0.000} & 0.259 $\pm$ 0.012 & 0.476 $\pm$ 0.000 \\ 
SVM & 0.933 $\pm$ 0.002 & 0.976 $\pm$ 0.002 & 0.239 $\pm$ 0.009 & 0.538 $\pm$ 0.002 \\ 
MLP & 0.968 $\pm$ 0.002 & 0.993 $\pm$ 0.001 & 0.227 $\pm$ 0.011 & 0.451 $\pm$ 0.002 \\ 
\textbf{EFC} & 0.916 $\pm$ 0.002 & 0.981 $\pm$ 0.001 & \textbf{0.641 $\pm$ 0.002} & 0.664 $\pm$ 0.002 \\ 
\hline
Ensemble \\
\hline
AB & 0.995 $\pm$ 0.001 & 1.000 $\pm$ 0.000 & 0.270 $\pm$ 0.013 & 0.660 $\pm$ 0.001 \\ 
RF & 0.997 $\pm$ 0.000 & 1.000 $\pm$ 0.000 & 0.089 $\pm$ 0.032 & 0.623 $\pm$ 0.000 \\ 
\hline
\end{tabular}
\end{adjustbox}
\end{table}

Taken as a whole, the results presented in this subsection show that EFC is better at adapting to other domains than classical ML-based classifiers on average.
In addition to that, it is possible to see that EFC achieves AUC values similar to the best ML algorithms when trained and tested in the same domain, showing that it is capable of performing well even if trained with only half of the information (benign data only) when compared to other classifiers (using malicious and benign data).
Not using malicious samples in the training phase is likely to be the reason why EFC is so good at adapting to other domains.
EFC's increased capability for domain adaptation when there is a significant difference in data distribution is a highly desirable trait in network flow-based classifiers, since changes in traffic composition are expected to be very frequent, and new kinds of attacks are generated continuously.

Finally, we believe EFC to be an useful tool for network managers, given (i) its more realistic requirements for training (only benign traffic that can be easily captured in the target network), (ii) its adaptability when faced with changes in traffic patterns, and (iii) the possibility to identify which flow features cause a specific network flow to be classified as benign or malicious.
However, there is still great room for improvement.
One possibility would be to incorporate EFC as a first step of a two-step NIDS, in which the flow samples detected as malicious by EFC would be then sent to deep package inspection.
Another possibility is to implement a dynamic threshold that would be adaptable to different network situations, improving classification accuracy.
There is also the possibility of performing feature selection previous to model inference, which would greatly reduce the time spent in the model inference phase and possibly also improve classification accuracy.
And finally, it would be possible to implement EFC to perform traffic classification at different points in a distributed network.
In the following, we present our conclusions and future work directions.

\section{Conclusion} \label{sec:conclusions}
In this work, we present a new flow-based classifier for network intrusion detection called Energy-based Flow Classifier (EFC).
In the training phase, EFC infers a statistical model based solely on benign traffic samples.
Afterward, this statistical model is used to classify network flows in benign or malicious based on "energy" values.
Our results show that EFC is capable of correctly performing network flow binary classification considering three different datasets.
F1 score (around 97\% at best) and AUC (around 99\% at best) values obtained using EFC are comparable to the values obtained with other classical ML-based classifiers, such as k-nearest neighbors, decision tree and multilayer perceptron, even though EFC uses only half of the information in the training phase compared to the other algorithms.

In addition to that, we analyzed different classifiers in terms of their capability for domain adaptation and observed that EFC is more suitable to that than classical ML-based algorithms.
In all the experiments performed to evaluate that over different datasets, EFC outperformed the other classifiers in F1-score and was among the best ones in AUC.
We understand that EFC's capability for domain adaptation is probably linked to the fact that the model inference is based only on benign samples, which helps preventing overfit.

Considering the advantages presented, we believe EFC to be a promising algorithm to perform flow-based traffic classification.
Nevertheless, despite the promising results achieved, there is still room for further testing and improvement.
In future work, we aim at performing a more comprehensive investigation of EFC's applicability to real-world data and different contexts, such as fraud analysis in bank data.
We are already working in a multiclass version of EFC that will be capable of identifying different kinds of known attacks, as well as unknown suspicious flow samples.
Finally, we will investigate which improvements can be attained by using a dynamic threshold in EFC and performing feature selection previous to model inference.

\section*{Acknowledgment}
The authors would like to thank Luís Paulo Faina Garcia for helping with dataset analysis. Matt Bishop was supported by the National Science Foundation under Grant Number OAC-1739025. Any opinions, findings, and conclusions or recommendations expressed in this material are those of the author(s) and do not necessarily reflect the views of the National Science Foundation.
João Gondim gratefully acknowledges the support from Project "EAGER: USBRCCR: Collaborative: Securing Networks in the Programmable Data Plane Era" funded by NSF (National Science Foundation), RNP (Brazilian National Research Network) and GigaCandanga.
\ifCLASSOPTIONcaptionsoff
  \newpage
\fi

\bibliographystyle{IEEEtran}
\bibliography{bare_jrnl.bib}

\hyphenation{CIC/UnB}

\begin{IEEEbiography}[{\includegraphics[width=1in,height=1.25in,clip,keepaspectratio]{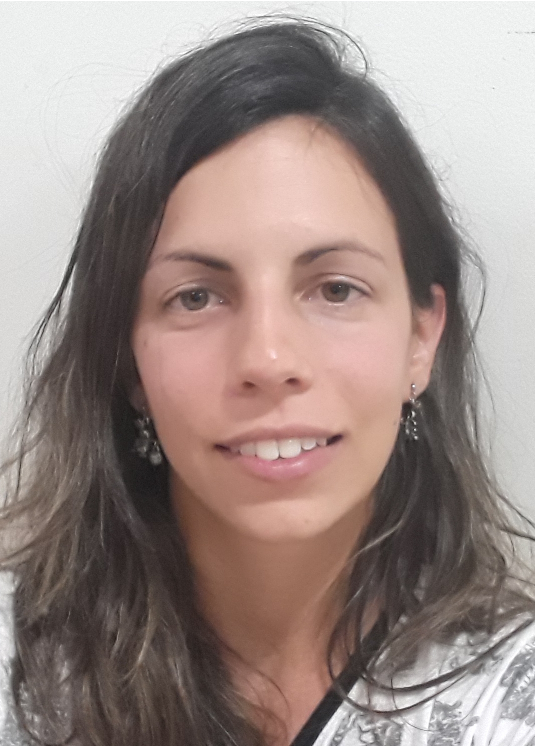}}]{Camila F. T. Pontes}

is a student at the University of Brasilia (UnB), Brasilia, DF, Brazil. She received her M.Sc. degree in Molecular Biology in 2016 from UnB and is currently an undergrad student at the Department of Computer Science (CIC/UnB). Her research interests are Computational and Theoretical Biology and Network Security.
\end{IEEEbiography}

\begin{IEEEbiography}[{\includegraphics[width=1in,height=1.25in,clip,keepaspectratio]{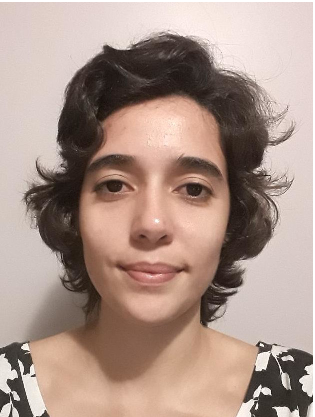}}]{Manuela M. C. de Souza} is an undergrad Computer Science student at University of Brasilia (UnB), Brasilia, DF, Brazil.
Her research interest is Network Security.
\end{IEEEbiography}

\begin{IEEEbiography}[{\includegraphics[width=1in,height=1.25in,clip,keepaspectratio]{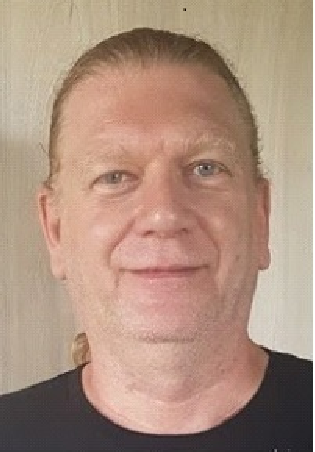}}]{Jo\~ao J. C. Gondim} was  awarded an M.Sc. in Computing Science at Imperial College, University of London, in 1987 and a Ph.D. in Electrical Engineering at UnB (University of Brasilia, 2017). He is an adjunct professor at Department of Computing Science (CIC) at UnB where he is a tenured member of faculty. His research interests are network, information and cyber security.
\end{IEEEbiography}

\begin{IEEEbiography}[{\includegraphics[width=1in,height=1.25in,clip,keepaspectratio]{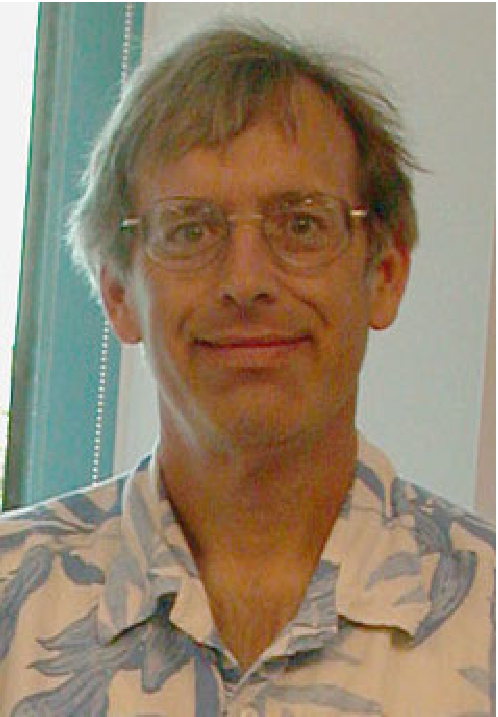}}]{Matt Bishop}
received his Ph.D. in computer science from Purdue University, where he specialized in computer security, in 1984. His main research area is the analysis of vulnerabilities in computer systems. The second edition of his textbook, Computer Security: Art and Science, was published in 2002 by Addison-Wesley Professional. He is currently a co-director of the Computer Security Laboratory at the University of California Davis.
\end{IEEEbiography}

\begin{IEEEbiography}
    [{\includegraphics[width=1in,height=1.25in,keepaspectratio]{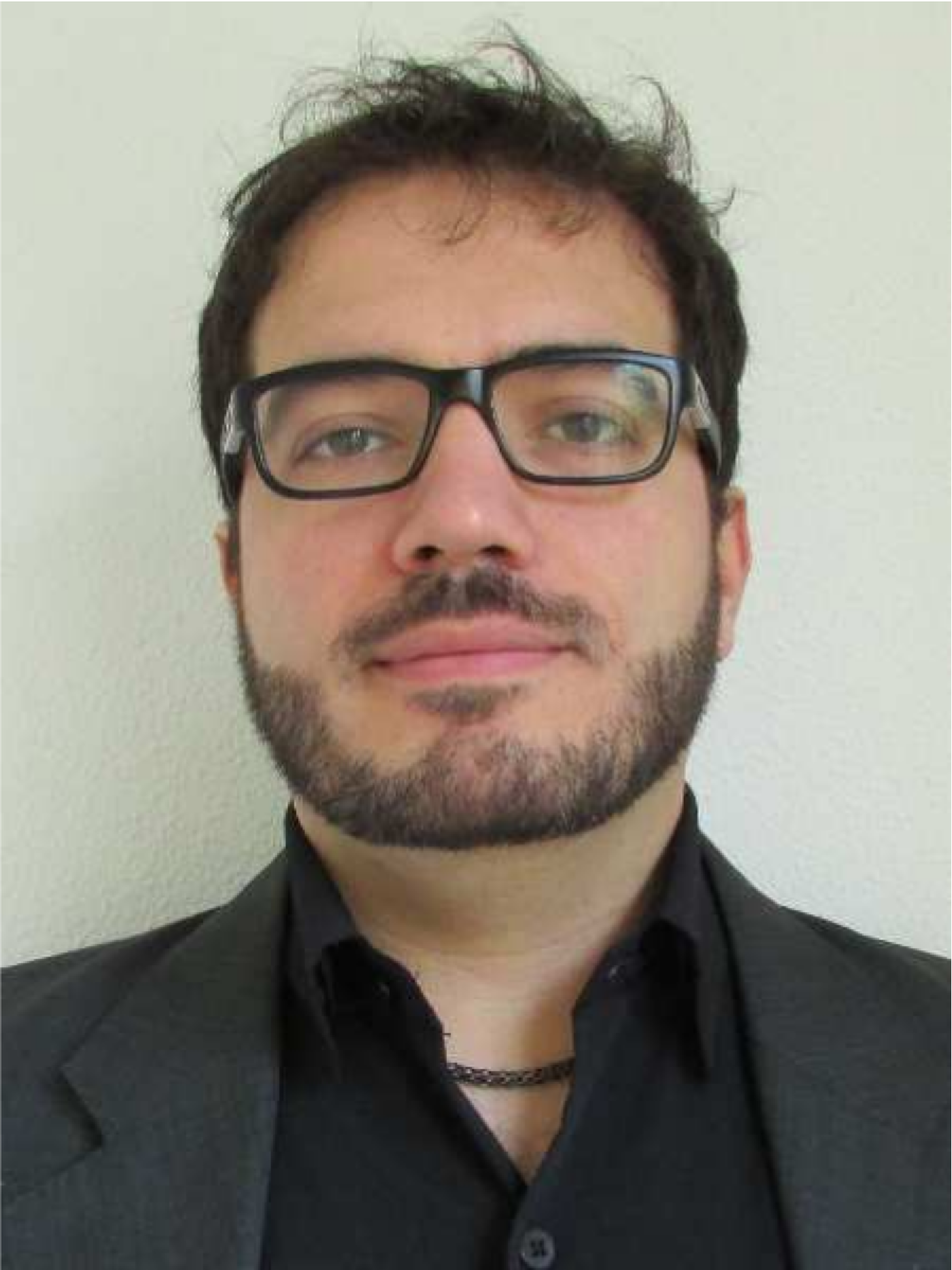}}]{Marcelo Antonio Marotta} 
is an adjunct professor at the University of Brasilia, Brasilia, DF, Brazil. He received his Ph.D. degree in Computer Science in 2019 from the Institute of Informatics (INF) of the Federal University of Rio Grande do Sul (UFRGS), Brazil. His research involves Heterogeneous Cloud Radio Access Networks, Internet of Things, Software Defined Radio, Cognitive Radio Networks, and Network Security.
\end{IEEEbiography}

\vfill

\end{document}